\documentstyle[12pt]{article}
\def\hybrid{\topmargin 0pt      \oddsidemargin 0pt
	\headheight 0pt \headsep 0pt
	\textheight 9in         
	\textwidth 6.25in       
	\marginparwidth .875in
	\parskip 5pt plus 1pt   \jot = 1.5ex}

\catcode`\@=11
\def\marginnote#1{}
\newcount\hour
\newcount\minute
\newtoks\amorpm
\hour=\time\divide\hour by60
\minute=\time{\multiply\hour by60 \global\advance\minute by-\hour}
\edef\standardtime{{\ifnum\hour<12 \global\amorpm={am}%
	\else\global\amorpm={pm}\advance\hour by-12 \fi
	\ifnum\hour=0 \hour=12 \fi
	\number\hour:\ifnum\minute<10 0\fi\number\minute\the\amorpm}}
\edef\militarytime{\number\hour:\ifnum\minute<10 0\fi\number\minute}

\def\draftlabel#1{{\@bsphack\if@filesw {\let\thepage\relax
   \xdef\@gtempa{\write\@auxout{\string
      \newlabel{#1}{{\@currentlabel}{\thepage}}}}}\@gtempa
   \if@nobreak \ifvmode\nobreak\fi\fi\fi\@esphack}
	\gdef\@eqnlabel{#1}}
\def\@eqnlabel{}
\def\@vacuum{}
\def\draftmarginnote#1{\marginpar{\raggedright\scriptsize\tt#1}}

\def\draft{\oddsidemargin -.5truein
	\def\@oddfoot{\sl preliminary draft \hfil
	\rm\thepage\hfil\sl\today\quad\militarytime}
	\let\@evenfoot\@oddfoot \overfullrule 3pt
	\let\label=\draftlabel
	\let\marginnote=\draftmarginnote
   \def\@eqnnum{(\theequation)\rlap{\kern\marginparsep\tt\@eqnlabel}%
\global\let\@eqnlabel\@vacuum}  }

\def\numberbysection{\@addtoreset{equation}{section}
	\def\theequation{\thesection.\arabic{equation}}}

\def\underline#1{\relax\ifmmode\@@underline#1\else
	$\@@underline{\hbox{#1}}$\relax\fi}

\def\titlepage{\@restonecolfalse\if@twocolumn\@restonecoltrue\onecolumn
     \else \newpage \fi \thispagestyle{empty}\c@page\z@
	\def\thefootnote{\fnsymbol{footnote}} }

\def\endtitlepage{\if@restonecol\twocolumn \else  \fi
	\def\thefootnote{\arabic{footnote}}
	\setcounter{footnote}{0}}  
\catcode`@=12
\relax

\def\beq{\begin{equation}}
\def\eeq{\end{equation}}
\def\bea{\begin{eqnarray}}
\def\eea{\end{eqnarray}}
\def\bar{\overline}

\def\nn{\nonumber}

\relax
\hybrid
\begin{document}
\begin{titlepage}
\setcounter{page}{0}
\begin{center}
 \hfill   CERN--TH.6502/92 \\
 \hfill   PAR--LPTHE 92--17\\[.4in]
{\large THE OPERATOR ALGEBRA OF THE DISCRETE STATE OPERATORS
 IN 2D GRAVITY WITH NON-VANISHING COSMOLOGICAL CONSTANT}\\[.4in]
	\large   Vl.S.Dotsenko\footnote{E-mail address:
	dotsenko@lpthe.jussieu.fr ; dotsenko@cernvm.bitnet}
\footnote{Permanent
address: Landau Institute for Theoretical
Physics, Moscow.}\\[.2in]
	\normalsize {\it Theory Division, CERN \\
	CH--1211 Geneva 23, Switzerland \\
	 and \\
LPTHE\/}\footnote{Laboratoire associ\'e No. 280 au
CNRS}\\
       \it  Universit\'e Pierre et Marie Curie, PARIS VI\\
	Tour 16, 1$^{\it er}$ \'etage \\
	4 place Jussieu\\
	75252 Paris CEDEX 05, FRANCE\\

\end{center}

\vskip .3in
\centerline{ ABSTRACT}
\begin{quotation}

Remarks are given to the structure of physical states
in the 2D gravity coupled to $C\leq 1$ matter.
The operator algebra of the discrete state operators
is calculated for the theory with non-vanishing cosmological
constant.

\end{quotation}
\vskip 1cm
{\flushleft CERN--TH.6502/92 \\
May, 1992}
\end{titlepage}
\newpage

\section {Introduction and results}
\numberbysection
The extra discrete physical states in the 2D gravity coupled
to matter have been a subject of a number of recent studies.
Their existance had been observed in the matrix
model approach \cite{a1} for $C= 1$ theory, and they have
been defined in the continuum theory and further
analyzed in \cite{a2,a3,a4,a5,a6,a7}, for more general
class of theories.

The operator algebra of the operators corresponding to these
extra discrete states have been defined in \cite{a8,a9},
in the framework of $C=1$ theory, for the vanishing value
of the cosmological constant. It was found that the chiral
current type operators form the $w_{\infty}$ algebra,
while the zero-form, zero ghost number operators form
the ground ring algebra. In the matrix model approach
the related results have been obtained in the papers \cite{a10}.

Further consequences of these algebras, in the form of Ward
identities for amplitudes of tachyonic operators have been analyzed
recently in \cite{a11,a12,a13,a131,a132}.

Here we shall calculate the deformation of the above mentioned
algebras by the presence in the theory of a nonvanishing
cosmological constant. Work in that direction have been done
in a number of recent papers \cite{a14,a15,a16,a17}.

But we shall repeat first some of the remarks made in \cite{a17}
to the structure of the physical states in 2D gravity coupled
to the minimal model as matter. This is to make further
calculation of the algebras applicable to a wider class
of theories.

2D gravity coupled to minimal matter could be represented,
apart from ghosts, by two free fields with background charges,
with the stress-energy tensor
\bea
&&T=T_{M}+T_{L}
\label{L1}\\
&&T_{M}=-\frac{1}{2}(\partial\varphi_{M})^{2}
+i\alpha_{0}\partial^{2}\varphi_{M}, \quad
T_{L}=-\frac{1}{2}(\partial\varphi_{L})^{2}
+\beta_{0}\partial^{2}\varphi_{L}
\label{L2}
\eea
Here $\varphi_{M}$ and $\varphi_{L}$ are normalized as:
\beq
\langle \varphi_{M}\varphi_{M} \rangle=
\langle \varphi_{L}\varphi_{L} \rangle=
\log \frac{1}{|z-z'|^{2}}
\label{L3}
\eeq
The field $\varphi_{M}$ is to represent, by the Coulomb gas
technique, the conformal theory of matter, and $\varphi_{L}$
that of Liouville. The relation for the central charges is:
\bea
&&C=C_{M}+C_{L}=26
\label{L4}\\
&&C_{M}=1-12\alpha^{2}_{0}, \quad
C_{L}=1+12\beta^{2}_{0}, \quad
\beta^{2}_{0}-\alpha^{2}_{0}=2
\label{L5}
\eea

The usual physical state operators will be of the form,
by David and Distler-Kawai arguments \cite{a18}:
\bea
&&\Phi^{(\mp)}_{n'.n}(z,\bar{z})
=exp(i\alpha_{n'.n}\varphi_{M}(z,\bar{z})
+\beta^{\mp}_{n'.n}\varphi_{L}(z,\bar{z}))
\label{L6}\\
&&\alpha_{n'.n}
=\frac{1-n'}{2}\alpha_{-}
+\frac{1-n}{2}\alpha_{+}
\label{L7}\\
&&\beta_{n'.n}
=\frac{1-n'}{2}\beta_{-}
+\frac{1-n}{2}\beta_{+}, \quad
\beta^{\mp}_{n'.n}=\beta_{-n'.n},\beta_{n'.-n}
\label{L8}\\
&&\alpha_{\pm}=\alpha_{0}\pm \sqrt{\alpha_{0}^{2}+2}, \quad
\beta_{\pm}=\beta_{0}\pm \sqrt{\beta_{0}^{2}-2}
=\pm \alpha_{\pm}
\label{L9}
\eea
with $(n'.n)$ taking their values inside the basic
conformal grid:
\beq
1\leq n' \leq p'-1, \quad 1\leq n \leq p-1
\label{L10}
\eeq
Here $p',p$ : $\alpha_{+}^{2}=2p'/p,\quad p',p$ being relative
prime numbers.
The unusual physical operators, the extra ones as compared
to the minimal midel without gravity, are the special
states in the modules of the states $\alpha_{n'.n}$ in
(\ref{L6}) (with $\beta^{\mp}_{n'.n}$ to be shifted
appropriately), or the states (\ref{L6}) themselves
with $(n'.n)$ outside the basic grid (\ref{L10}).

Under the linear transformation in the system of two
fields $\varphi_{M}$ and $\varphi_{L}$:
\bea
\varphi_{M}(z,\bar{z})
=\frac{1}{2}(-i\alpha_{0}\varphi_{1}+\beta_{0}\varphi_{2})
\nn\\
\varphi_{L}(z,\bar{z})
=\frac{1}{2}(-\beta_{0}\varphi_{1}-i\alpha_{0}\varphi_{2})
\label{L11}
\eea
the stress-energy tensor (\ref{L1}) takes the form
\beq
T=-\frac{1}{4}(\partial\varphi_{2})^{2}
-\frac{1}{4}(\partial\varphi_{1})^{2}
-\partial^{2}\varphi_{1}=T_{2}+T_{1}
\label{L12}
\eeq
with fields $\varphi_{1}$ and $\varphi_{2}$ normalized as:
\beq
\langle\varphi_{1}\varphi_{1}\rangle
=\langle\varphi_{2}\varphi_{2}\rangle
=2\log\frac{1}{|z-z'|^{2}}
\label{L13}
\eeq
By (12) we observe that we get an effective $C=1$ theory:
$\varphi_{2}$-`matter', with $\alpha_{0}=0$, $C_{2}=1$,
and $\varphi_{1}$-`Liouville', with $\beta_{0}=-1$,
$C_{1}=25$.

The operators in (\ref{L6}) become:
\bea
&&\Phi^{(-)}_{n'.n}(z,\bar{z})=
exp(ij^{(-)}_{n'.n}\varphi_{2}+(-1-j^{(-)}_{n'.n})\varphi_{1})
\nn\\
&&j^{(-)}_{n'.n}=\frac{n'}{2}\rho'-\frac{n}{2}, \quad
\rho'=\frac{(\alpha)^{2}_{-}}{2}=\frac{p}{p'}
\label{L14}\\
&&\Phi^{(+)}_{n'.n}(z,\bar{z})=
exp(ij^{(+)}_{n'.n}\varphi_{2}+(-1+j^{(+)}_{n'.n})\varphi_{1})
\nn\\
&&j^{(+)}_{n'.n}=\frac{n'}{2}-\frac{n}{2}\rho, \quad
\rho=\frac{(\alpha)^{2}_{+}}{2}=\frac{p'}{p}
\label{L15}
\eea

Taking e.g. the branch (\ref{L15}) we observe that the states
$(n'.0)$ are mapped, by the linear transformation (\ref{L11}),
onto the $su(2)$ discrete tachyonic states of the $C=1$ theory.
Their modules involve $su(2)$ multiplets of extra physical
states \cite{a2,a3,a5}. The usual physical states are mapped
onto the tachyonic states with fractional values of
the $su(2)$ momenta.

Notice that the extra physical states in the modules
of the $\alpha_{n'.n}$ states (\ref{L6}) (with
$\beta^{\mp}_{n'.n}$ to be shifted)
get assembled in the modules of the $su(2)$ tachyonic
states, $(n'.0)$ or $(0.n)$, after the linear transformation
(\ref{L11}). We have to check for the first extra state
in the module of $\alpha_{n'.n}$, as the inclusion repeats
itself. The first extra state will have the form:
\beq
(...) \exp(i\alpha_{n'.n}\varphi_{M}(z,\bar{z})
+\beta_{n'.n}\varphi_{L}(z,\bar{z}))
\label{L16}
\eeq
$\beta_{n',-n}$ is replaced with $\beta_{n'.n}$, and the conformal
dimension of the exponential operator becomes $1-n'n$;
$(...)$ stands for a polynomial of $\varphi_{M}$, $\varphi_{L}$
oscillator states, of level $n'n$. The exponential in (\ref{L16})
transforms, under (\ref{L11}), into
\beq
(...)\exp(i\frac{n'-n}{2}\varphi_{2}
+(-1+\frac{n'+n}{2})\varphi_{1})
\label{L17}
\eeq
This is the state in the module of $j=\frac{n'-n}{2}$
`matter' state.

After these remarks it is clear that we may as well
stay in the framework of the $C=1$ theory with gravity,
when calculating amplitudes (correlation functions).
In this respect it is more a question of what particular
set of operators we are interested in, than what is
the $C$-matter of the theory.

We shall give next the  results for the operator algebra
of $su(2)$ physical state operators \cite{a8}:
\beq
{\cal T}_{j,m}=T_{j,m}\bar{T}_{j,m}, \quad
{\cal O}_{j,m}=O_{j,m}\bar{O}_{j,m}
\label{L18}
\eeq
and
\beq
{\cal W}_{j,m}=T_{j+1,m}\bar{O}_{j,m}
\label{L19}
\eeq
$T_{j,m}$ are the conformal dimension one chiral operators:
\bea
&&T_{j,m}=(H^{-})^{j-m}\exp(ijX+(-1+j)\phi)
\label{L20}\\
&&H^{-}=\oint\frac{du}{2\pi i}\exp(-iX(u))
\label{L21}
\eea
They form $su(2)$ multiplets. The importance of the chiral
(antichiral) operators $O_{j,m}$ ($\bar{O}_{j,m}$) have been
stressed in \cite{a8}. The $j=1/2$ states have the following
explicit form:
\bea
&&O_{\frac{1}{2},\frac{1}{2}}=
:(cb+\frac{i}{2}\partial X-\frac{1}{2}\partial\phi)
\exp(\frac{i}{2}X+\frac{1}{2}\phi):
\label{L22}\\
&&O_{\frac{1}{2},-\frac{1}{2}}=H^{-}O_{\frac{1}{2},\frac{1}{2}}=
:(cb-\frac{i}{2}\partial X-\frac{1}{2}\partial\phi)
\exp(-\frac{i}{2}X+\frac{1}{2}\phi):
\label{L23}
\eea
We have assumed in (\ref{L20}--\ref{L23}) the notations
and normalizations of \cite{a9,a11}. In particular
\bea
&&X=\sqrt{2}\varphi_{M},\quad \phi=\sqrt{2}\varphi_{L},
\quad \alpha_{0}=0,\quad \beta_{0}=-1
\nn\\
&&\langle XX \rangle = \langle \phi\phi \rangle =
2\log\frac{1}{|z-z'|^{2}}
\label{L24}
\eea
$b$, $c$ are covariant ghosts, dimension 2,--1 chiral fields.
In the fully bosonized form the operator
$O_{\frac{1}{2},\frac{1}{2}}$ in (\ref{L22}) becomes:
\beq
O_{\frac{1}{2},\frac{1}{2}}=
:(i\partial\varphi+\frac{i}{2}\partial X-\frac{1}{2}\partial\phi)
\exp(\frac{i}{2}X+\frac{1}{2}\phi):
\label{L22'}\\
\eeq
Here \cite{a19}
\bea
&&c=\exp(i\varphi),\quad b=\exp(-i\varphi),\quad :cb:=i\varphi
\nn\\
&&\langle\varphi\varphi\rangle=\log\frac{1}{(z-z')}
\label{L25}
\eea
The operator $\bar{O}_{j,m}$ in (\ref{L19}) balance
the left-right momenta of physical chiral operators,
the currents ${\cal W}_{j,m}$ \cite{a8}.

Let us introduce the following operator:
\bea
&&\delta=-\oint\frac{du}{2\pi i}b(u)
\exp(-\frac{i}{2}X+\frac{1}{2}\phi)
\nn\\
&&=-\oint\frac{du}{2\pi i}
\exp(-i\varphi-\frac{i}{2}X+\frac{1}{2}\phi)
\label{L26}
\eea
It is easy to check that
\bea
&&\delta(c(z)T_{1,1}(z))
\nn\\
&&=-\oint\frac{du}{2\pi i}
\exp(-i\varphi-\frac{i}{2}X+\frac{1}{2}\phi)
\times \exp(i\phi(z)+iX(z))=O_{\frac{1}{2},\frac{1}{2}}(z)
\label{L27}
\eea
Easy to see also that $\delta^{2}=0$, $\delta$ commutes
with $H^{-}$, eq.(\ref{L21}), and it can be checked,
with some algebra, that $\delta$ commutes with the BRST
operator
\beq
d=\oint\frac{du}{2\pi i}:c(T_{M}+T_{L}+\frac{1}{2}T_{gh})
\label{L28}
\eeq
-- this is up to some bare null states, which are supposed
to decouple in the amplitudes.
Then, in the general case, we will have, by matching the quantum
numbers:
\beq
O_{j,m}=\delta(cT_{j+\frac{1}{2},m+\frac{1}{2}})
\label{L29}
\eeq

The calculation of each particular term of the operator
algebra factorizes onto chiral and antichiral factor
calculations, up to a small detail for the Liouville part
which is remarked on in Section 2. The calculation of
the operator algebra of the $T_{j,m}$ chiral factors in
(\ref{L18}),(\ref{L19}) has already been done in \cite{a17},
and is reproduced in Section 2, in a slightly different way.
The calculation of the algebra of the chiral operators
$O_{j,m}$ can be done using the representation (\ref{L29}).
This is done in Section 3.

The general form of the algebras could be written as
follows:
\bea
&&{\cal T}_{j_{1},m_{1}}(z,\bar{z})\times
{\cal T}_{j_{2},m_{2}}(z',\bar{z}')=
\frac{1}{|z-z'|^{2}}\sum_{j_{3}}
D^{({\cal T})}_{(j_{1},m_{1}),(j_{2},m_{2}),(j_{3},m_{3})}
{\cal T}_{j_{3},m_{3}}(z',\bar{z}')
\label{L30}\\
&&{\cal O}_{j_{1},m_{1}}(z,\bar{z})\times
{\cal O}_{j_{2},m_{2}}(z',\bar{z}')=
\sum_{j_{3}}
D^{({\cal O})}_{(j_{1},m_{1}),(j_{2},m_{2}),(j_{3},m_{3})}
{\cal O}_{j_{3},m_{3}}(z',\bar{z}')
\label{L31}\\
&&{\cal W}_{j_{1},m_{1}}(z,\bar{z})\times
{\cal W}_{j_{2},m_{2}}(z',\bar{z}')=
\frac{1}{(z-z')}\sum_{j_{3}}
D^{({\cal W})}_{(j_{1},m_{1}),(j_{2},m_{2}),(j_{3},m_{3})}
{\cal W}_{j_{3},m_{3}}(z',\bar{z}')
\label{L32}
\eea
with the operator algebra structure constants having the general
form:
\bea
&&D^{({\cal T})}_{(j_{1},m_{1}),(j_{2},m_{2}),(j_{3},m_{3})}
=(A_{(j_{1},m_{1}),(j_{2},m_{2}),(j_{3},m_{3})})^{2}
\label{L33}\\
&&D^{({\cal O})}_{(j_{1},m_{1}),(j_{2},m_{2}),(j_{3},m_{3})}
=(B_{(j_{1},m_{1}),(j_{2},m_{2}),(j_{3},m_{3})})^{2}
\label{L34}\\
&&D^{({\cal W})}_{(j_{1},m_{1}),(j_{2},m_{2}),(j_{3},m_{3})}=
A_{(j_{1}+1,m_{1}),(j_{2}+1,m_{2}),(j_{3}+1,m_{3})}
B_{(j_{1},m_{1}),(j_{2},m_{2}),(j_{3},m_{3})}
\label{L35}
\eea
In the r.h.s. of (\ref{L30}--\ref{L32}) just the BRST
cohomology states are kept. The component constants
$A$ and $B$ are found, in Section 2 and Section 3,
to have the following form,
for the appropriately normalized operators:
\bea
&&A_{(j_{1},m_{1}),(j_{2},m_{2}),(j_{3},m_{3})}=
(^{j_{1}\quad\,\,\, j_{2}\quad\,\,\, j_{3}}_{m_{1}\quad
m_{2}\quad -m_{3}})
d^{(A)}_{j_{1},j_{2},j_{3}}
\label{L36}\\
&&d^{(A)}_{j_{1},j_{2},j_{3}}=
\frac{(j_{1}+j_{2}-j_{3})!(j_{1}-j_{2}+j_{3})!(-j_{1}
+j_{2}+j_{3})!}{\Delta^{1/2}_{j_{1},j_{2},j_{3}}[(2j_{1}-1)!
(2j_{2}-1)!(2j_{3}-1)!]^{1/2}}
\label{L37}\\
&&B_{(j_{1},m_{1}),(j_{2},m_{2}),(j_{3},m_{3})}=
(^{j_{1}\quad\,\,\, j_{2}\quad\,\,\, j_{3}}_{m_{1}\quad
m_{2}\quad -m_{3}})
d^{(B)}_{j_{1},j_{2},j_{3}}
\label{L38}\\
&&d^{(B)}_{j_{1},j_{2},j_{3}}=
\frac{(j_{1}+j_{2}-j_{3})!(j_{1}-j_{2}+j_{3})!(-j_{1}
+j_{2}+j_{3})!}{\Delta^{1/2}_{j_{1},j_{2},j_{3}}[(2j_{1})!
(2j_{2})!(2j_{3})!]^{1/2}}
\label{L39}
\eea
Here
$(^{j_{1}\quad\,\,\, j_{2}\quad\,\,\, j_{3}}_{m_{1}\quad
m_{2}\quad m_{3}})$
is the $su(2)$ $3j$ symbol. The factor
$\Delta^{1/2}_{j_{1},j_{2},j_{3}}$, which in fact cancels
in the products of the $3j$ symbols and the $d$ coefficients
in (\ref{L36}),(\ref{L38}), is given by
\beq
\Delta^{1/2}_{j_{1},j_{2},j_{3}}=
[\frac{(j_{1}+j_{2}-j_{3})!(j_{1}-j_{2}+j_{3})!(-j_{1}
+j_{2}+j_{3})!}{(j_{1}+j_{2}+j_{3}+1)!}]^{1/2}
\label{L40}
\eeq

The structure constants $D^{({\cal W})}$ of the algebra
of currents (\ref{L32}) can be written more explicitly
as
\beq
D^{({\cal W})}_{(j_{1},m_{1}),(j_{2},m_{2}),(j_{3},m_{3})}=
(^{j_{1}+1\quad\,\, j_{2}+1\quad\,\, j_{3}+1}_{m_{1}\quad\quad
m_{2}\quad\quad -m_{3}})
(^{j_{1}\quad\,\,\, j_{2}\quad\,\,\, j_{3}}_{m_{1}\quad
m_{2}\quad -m_{3}})
d^{(A)}_{j_{1}+1,j_{2}+1,j_{3}+1}
d^{(B)}_{j_{1},j_{2},j_{3}}
\label{L41}
\eeq
Under the permutation of e.g. $(j_{1},m_{1})$ and
$(j_{2},m_{2})$ the $3j$ symbols produce the factors:
\beq
(-1)^{j_{1}+j_{2}+j_{3}+3}\times (-1)^{j_{1}+j_{2}+j_{3}}=-1
\label{L42}
\eeq
So the constants (\ref{L41}) are properly antisymmetric.
The current constants should satisfy also the Jacobi identities,
which is harder to verify. This has not been done yet.

\section {Calculation of the operator algebra of $T_{j,m}$}

We shall look at the product of two particular operators
and shall calculate the coefficient at a particular term
(operator) in the r.h.s.:
\beq
T_{j_{1},j_{1}}\times T_{j_{2},j_{3}-j_{1}}
\rightarrow ...\quad T_{j_{3},j_{3}}
\label{L43}
\eeq
The operators $T_{j,m}$ are defined in (\ref{L20}).
It is easy to check that the coefficient at $T_{j_{3},j_{3}}$
is given by:
\bea
&&I=I_{M}I_{L}
\label{L44}\\
&&I_{M}=\langle\exp(ij_{1}X(0)\times
(\prod^{k}_{1}\oint\frac{dv_{i}}{2\pi i}\exp(-iX(v_{i})))
\times
\exp(ij_{2}X(1))\rangle
\nn\\
&&=\prod^{k}_{1}\oint\frac{dv_{i}}{2\pi i}
(v_{i})^{-2j_{1}}(v_{i}-1)^{-2j_{2}}
\prod^{k}_{i<j}(v_{i}-v_{j})^{2}
\label{L45}\\
&&=k!\prod^{k}_{1}\frac{\Gamma(i)\Gamma(2j_{3}
+i)}{\Gamma(2j_{1}+1-i)\Gamma(2j_{2}+1-i)}
\label{L46}
\eea
(double use of $i$ in (\ref{L45}) should not be confusing).
Here $k\equiv k_{M}=j_{1}+j_{2}-j_{3}$. The result (\ref{L46})
is obtained by transforming the contours in (\ref{L45}),
which encircle the point $z=1$, to the standart ones in \cite{a20},
and then substituting the result for the integral.
Some further slight transformations of the resulting
products has been done to put the result in a more symmetric form.
\bea
&&I_{L}=\frac{(\mu)^{k_{L}}}{k_{L}!}
\langle\exp((-1+j_{1})\phi(0)\times
(\prod^{k_{L}}_{1}\int^{+\infty}_{-\infty} dw_{i}
\exp(-\phi(w_{i})))
\times
\exp((-1+j_{2})\phi(1))\rangle
\nn\\
&&=\frac{(\mu)^{k_{L}}}{k_{L}!}
\prod^{k_{L}}_{1}\int^{+\infty}_{-\infty} dw_{i}
(w_{i})^{2(-1+j_{1})}(w_{i}-1)^{2(-1+j_{2})}
\prod^{k_{L}}_{i<j}(w_{i}-w_{j})^{-2}
\label{L47}\\
&&=(\frac{\mu}{\Gamma(-1)})^{k_{L}}
\prod^{k_{L}}_{1}\frac{\Gamma(2j_{1}-i)\Gamma(2j_{2}
-i)}{\Gamma(1+i)\Gamma(2j_{3}+1+i)}
\label{L48}
\eea
Here $k_{L}=j_{1}+j_{2}-j_{3}-1=k_{M}-1$. The contours in
(\ref{L47}) extend from $-\infty$ to $+\infty$ going below
$z=0$ and above $z=1$. They had been deformed to the standard
ones, and the result for the integral in \cite{a20} had been
used.

For the product of $I_{M}I_{L}$ in (\ref{L44}) we get
\beq
I=(\frac{\mu}{\Gamma(-1)})^{k_{L}}(j_{1}+j_{2}-j_{3})!
\frac{\Gamma(2j_{3}+1)}{\Gamma(2j_{1})\Gamma(2j_{2})}
\label{L49}
\eeq

The remark is in order. The Liouville integral in (\ref{L47})
is presumed to come from the expansion over the term
\beq
\mu\int d^{2}w\exp(-\phi(w,\bar{w}))
\label{L50}
\eeq
in the action for the field $\phi$. The $k_{L}$th order term
is
\beq
\frac{(\mu)^{k_{L}}}{k_{L}!}\prod^{k_{L}}_{1}\int d^{2}w_{i}
\exp(-\phi(w_{i},\bar{w}_{i}))
\label{L51}
\eeq
It is being picked up by the (anomalous) conservation
of the $\phi$ field momenta (or `charges', for the theory
with the background charge $\beta_{0}=-1$). The integrals
in (\ref{L51}), which are over the 2D plane, are to be factorized
onto chiral and antichiral contour integrals, as is explained
e.g. in \cite{a21}. In the expansion over the exponential
term (\ref{L50}) the chiral and the antichiral integrals
will go in parallel, in equal numbers, so that the Liouville
momenta have to match on both sides of the operator product
algebra for the physical operators, i.e. in the relation
(\ref{L43}) supplied by the antichiral partners:
\beq
(T_{j_{1},j_{1}}\bar{T}_{j_{1},j_{1}})
\times (T_{j_{2},j_{3}-j_{1}}\bar{T}_{j_{2},j_{3}-j_{1}})
\rightarrow ...\quad (T_{j_{3},j_{3}}\bar{T}_{j_{3},j_{3}})
\label{L52}
\eeq

It should be noted also that, in undoing the 2D integrals
over the plane into the product of the contour integrals,
one gets, in a particular way of doing it, the contours
extending from $-\infty$ to $+\infty$, one contour going below
the other, and all going below the point $z=0$ and above
the point $z=1$. This is on the chiral side. In the corresponding
antichiral factor the integrations will be of the ordered type,
one point after another, in the interval $0<\bar{z}<1$.
This could be replaced also by the integral with all
the contours going from $\bar{z}=0$ to $\bar{z}=1$, one below
the other, which is to be devided then by $k_{L}!$.
This replacement is useful to avoid the formal problem
with the exponent $-2$ in (\ref{L47}). (Notice that
$j_{1}$, $j_{2}$ could be thought of, in the calculation
of the integral, as having general values, till the final result;
so the rest of the exponents in the integrals pose no problem).
Taken one way or another, the antichiral part will obviously
not supply the extra factor of $(\mu)^{k_{L}}$; but also
it will not produce the extra factor of
$(1/\Gamma(-1))^{k_{L}}$, as can be checked.
The factor
\beq
(\frac{\mu}{\Gamma(-1)})^{k_{L}}\equiv (\mu_{eff})^{k_{L}}
\label{L53}
\eeq
is in fact common for the chiral and the antichiral parts
of the operator product algebra.

Since we want to consider the $\mu$ deformed algebra,
the cosmological constant is being renormalized
in a singular way, as in (\ref{L53}),
which is the known point of the $C=1$ theory.
One also could think
of it as of a singular renormalization of the corresponding
screening operators, which would be the case for the $C<1$
theory rotated to the effective theory with $C=1$.

Performing the singular renormalization, or letting $\mu$
to vanish, appears in this respect as our own choice
of the theory, both being well defined.

By $su(2)$ invariance the coefficient of (\ref{L43})
in (\ref{L49}) should have the form:
\beq
(\mu_{eff})^{k_{L}}(^{j_{1}\quad\,\,\,j_{2}\quad\,\,\,
j_{3}}_{j_{1}\,\,\,\,\, j_{3}-j_{1}\,\,\,\,\, -j_{3}})
d_{j_{1},j_{2},j_{3}}
\label{L54}
\eeq
where the first factor is the $su(2)$ $3j$ symbol, which
in particular could be given as:
\bea
&&(^{j_{1}\quad\,\,\,j_{2}\quad\,\,\,j_{3}}_{m_{1}\quad
m_{2}\quad m_{3}})
\nn\\
&&=\Delta^{1/2}[(j_{1}+m_{1})!(j_{1}-m_{1})!
(j_{2}+m_{2})!(j_{2}-m_{2})!(j_{3}+m_{3})!(j_{3}-m_{3})!]^{1/2}
\nn\\
&&\times \sum_{i}(\frac{(-1)^{j_{1}-j_{2}-m_{3}+i}}{(i)!
(j_{1}+j_{2}-j_{3}-i)!(j_{1}-m_{1}-i)!(j_{2}+m_{2}-i)!}
\nn\\
&&\quad\quad\quad\quad\quad\quad\quad\quad
\frac{1}{(j_{3}-j_{2}+m_{1}+i)!(j_{3}-j_{1}-m_{2}+i)!})
\label{L55}
\eea
with $\Delta^{1/2}$ given in (\ref{L40}). Matching (\ref{L55})
with (\ref{L49}), to obtain the coefficients
$d_{j_{1},j_{2},j_{3}}$, we have to remove first the extra
$su(2)$ normalization of the operators $T_{j,m}$ in (\ref{L20}),
which is
\beq
\langle T_{j,m}^{+}\,T_{j,m} \rangle \propto
(2j)!\frac{(j-m)!}{(j+m)!}
\label{L56}
\eeq
Finally, for the coefficients of the operator product expansion
\beq
T_{j_{1},m_{1}}\times T_{j_{2},m_{2}} \rightarrow
\sum_{j_{3}}T_{j_{3},m_{3}}
\label{L57}
\eeq
we get
\beq
(^{j_{1}\quad\,\,\,j_{2}\quad\,\,\,j_{3}}_{m_{1}\quad
m_{2}\quad -m_{3}}) d_{j_{1},j_{2},j_{3}}
\label{L58}
\eeq
with
\bea
&&d_{j_{1},j_{2},j_{3}}=(\mu_{eff})^{j_{1}+j_{2}-j_{3}}
\frac{(j_{1}+j_{2}-j_{3})!(j_{1}-j_{2}+j_{3})!(-j_{1}
+j_{2}+j_{3})!}{\Delta^{1/2}_{j_{1},j_{2},j_{3}}[(2j_{1})!
(2j_{2})!(2j_{3})!]^{1/2}}
\nn\\
&&\quad\quad\quad\quad\quad\quad\quad\quad
\frac{(2j_{3})!}{(2j_{1}-1)!(2j_{2}-1)!}
\label{L59}
\eea
This could be given in a more symmetric form as:
\bea
&&(\mu_{eff})^{j_{1}+j_{2}-j_{3}}
\times\frac{(j_{1}+j_{2}-j_{3})!(j_{1}-j_{2}+j_{3})!(-j_{1}
+j_{2}+j_{3})!}{\Delta^{1/2}_{j_{1},j_{2},j_{3}}[(2j_{1}-1)!
(2j_{2}-1)!(2j_{3}-1)!]^{1/2}}
\nn\\
&&\quad\quad\quad\quad\quad\quad\quad\quad
\times [\frac{(2j_{3})}{(2j_{1})(2j_{2})}]^{1/2}
\times\frac{(2j_{3}-1)!}{(2j_{1}-1)!(2j_{2}-1)!}
\label{L60}
\eea

The above differs from the expression obtained in \cite{a17}
by the normalization factors of individual operators.
This is due to different representation of operators
used in \cite{a17} and here.

The extra factor of $1/\Gamma(0)$ in \cite{a17} is that
of the partition function normalization. It is absent here
since we are doing directly the operator product expansion,
instead of calculating the 3-point function.

The first, third, and fourth factors could be absorbed
into the normalization of the operators, which leaves us
with the symmetric coefficients $d^{(A)}_{j_{1},j_{2},j_{3}}$
in (\ref{L37}). We remark that we have not followed
and kept the overall sign factors, as they cancel eventually
in the products of the chiral and antichiral factors.

\section {Calculation of the operator algebra of $O_{j,m}$}

For the operators $O_{j,m}$, eq.(\ref{L29}), we shall again
calculate the coefficient in
\bea
O_{j_{1},j_{1}}\times O_{j_{2},j_{3}-j_{1}}
\rightarrow ...\quad O_{j_{3},j_{3}}
\label{L61}
\eea
which is easier, in a similar way as for the operators
$T_{j,m}$ in Section 2. We have:
\bea
&&O_{j_{1},j_{1}}(0)\times O_{j_{2},j_{3}-j_{1}}(1)
\nn\\
&&=\delta(c(0)T_{j_{1}+\frac{1}{2},j_{1}+\frac{1}{2}}(0))
\times \delta((H^{-})^{j_{2}-j_{3}+j_{1}}
c(1)T_{j_{2}+\frac{1}{2},j_{2}+\frac{1}{2}}(1))
\nn\\
&&\quad\quad\quad\quad\quad\quad\quad
\rightarrow ...\quad
\delta(cT_{j_{3}+\frac{1}{2},j_{3}+\frac{1}{2}})
\label{L62}
\eea
The contour of $\delta$ acting on the operator at $z=0$
in l.h.s. can be tranformed to the contour which
encircles both operators, at $z=0$ and at $z=1$,
which is appropriate to obtain eventually the operator
in r.h.s., plus the configuration where two $\delta$
contours encircle the operator at $z=1$. The last vanishes
because of $\delta^{2}=0$. Then it is sufficient to calculate
a slightly reduced operator algebra, that of
\beq
cT_{j_{1}+\frac{1}{2},j_{1}+\frac{1}{2}}
\times
\delta((H^{-})^{j_{2}-j_{3}+j_{1}}
cT_{j_{2}+\frac{1}{2},j_{2}+\frac{1}{2}})
\rightarrow ...\quad
cT_{j_{3}+\frac{1}{2},j_{3}+\frac{1}{2}}
\label{L63}
\eeq
It is easy to check that the coefficient at the operator
in r.h.s. of (\ref{L63}) is given by the integral:
\bea
&&I=
\langle
\exp(i(j_{1}+\frac{1}{2})X(0)+(-\frac{1}{2}+j_{1})\phi(0)
+i\varphi(0))
\nn\\
&&\times
\oint\frac{du}{2\pi i}\exp(-\frac{i}{2}X(u)+\frac{1}{2}\phi(u)
-i\varphi(0))\times\prod^{k_{M}}_{1}\oint\frac{dv_{i}}{2\pi i}
\exp(-iX(v_{i}))
\nn\\
&&\times \frac{1}{k_{L}!}\prod^{k_{L}}_{1}
\int^{+\infty}_{-\infty}dw_{i}\exp(-\phi(w_{i}))
\nn\\
&&\times
\exp(i(j_{2}+\frac{1}{2})X(1)+(-\frac{1}{2}+j_{2})\phi(1)
+i\varphi(1))\rangle
\nn\\
&&=
\oint\frac{du}{2\pi i}(u)^{-(2j_{1}+1)}(u-1)^{-(2j_{2}+1)}
\nn\\
&&\times
\prod^{k_{M}}_{1}\oint\frac{dv_{i}}{2\pi i}
(v_{i})^{-(2j_{1}+1)}(v_{i}-1)^{-(2j_{2}+1)}(v_{i}-u)
\prod^{k_{M}}_{i<j}(v_{i}-v_{j})^{2}
\nn\\
&&\times
\frac{1}{k_{L}!}\prod^{k_{L}}_{1}\int^{+\infty}_{-\infty}dw_{i}
(w_{i})^{2j_{1}-1}(w_{i}-1)^{2j_{i}-1}(w_{i}-u)
\prod^{k_{L}}_{i<j}(w_{i}-w_{j})^{-2}
\label{L64}
\eea
This could be put into the form:
\beq
I=\oint\frac{du}{2\pi i}(u)^{-(2j_{1}+1)}(u-1)^{-(2j_{2}+1)}
I_{M}(u)I_{L}(u)
\label{L65}
\eeq
with
\bea
&&I_{M}(u)=\prod^{k_{M}}_{1}\oint\frac{dv_{i}}{2\pi i}
(v_{i})^{-(2j_{1}+1)}(v_{i}-1)^{-(2j_{2}+1)}(v_{i}-u)
\prod^{k_{M}}_{i<j}(v_{i}-v_{j})^{2}
\label{L66}\\
&&I_{L}(u)=\frac{1}{k_{L}!}\prod^{k_{L}}_{1}\int^{+\infty}_{-\infty}
dw_{i}
(w_{i})^{2j_{1}-1}(w_{i}-1)^{2j_{i}-1}(w_{i}-u)
\prod^{k_{L}}_{i<j}(w_{i}-w_{j})^{-2}
\label{L67}
\eea
One checks that in this case (different from that in Sec.2)
\beq
k_{M}=k_{L}=j_{1}+j_{2}-j_{3}\equiv k
\label{L68}
\eeq
The $u$ contour in (\ref{L65}) encircles $z=1$, like
the $v_{i}$ contours in (\ref{L66}). The $w_{i}$ contours
go below $z=0$ and above $z=1$,-- see remarks on the Liouville
contours in Section 2. We dropped here the factor
$(\mu)^{k_{L}}$ in front of the Liouville integrals.

Some intermediate expressions are more symmetric if we
present the integrals (\ref{L66}),
(\ref{L67}) in the following forms:
\bea
&&I_{M}(u)\equiv (u)^{k}J_{M}(t), \quad t=1/u
\nn\\
&&J_{M}(t)=\prod^{k}_{1}\oint\frac{dv_{i}}{2\pi i}
(v_{i})^{-(2j_{1}+1)}(v_{i}-1)^{-(2j_{2}+1)}(v_{i}t-1)
\prod^{k}_{i<j}(v_{i}-v_{j})^{2}
\label{L69}\\
&&I_{L}(u)\equiv (u)^{k}J_{L}(t), \quad t=1/u
\nn\\
&&J_{L}(t)=\frac{1}{k!}\prod^{k}_{1}\int^{+\infty}_{-\infty}
dw_{i}
(w_{i})^{2j_{1}-1}(w_{i}-1)^{2j_{2}-1}(w_{i}t-1)
\prod^{k}_{i<j}(w_{i}-w_{j})^{-2}
\label{L70}
\eea
The integrals $J_{M}(t)$, $J_{L}(t)$ could be transformed
to the following forms:
\bea
&&J_{M}(t)=(\frac{\sin\pi(-2j_{2}-1)}{\pi})^{k}
\nn\\
&&\times\prod^{k}_{1}\int^{1}_{0}dx_{i}
(x_{i})^{2j_{3}+1}(1-x_{i})^{-2j_{2}-1}(t-x_{i})
\prod^{k}_{i<j}(x_{i}-x_{j})^{2}
\label{L71}\\
&&J_{L}(t)=(\frac{\sin\pi(2j_{2}-1)}{\pi})^{k}
\nn\\
&&\times\frac{1}{k!}\prod^{k}_{1}\int^{1}_{0}
dy_{i}
(y_{i})^{-3-2j_{3}}(1-y_{i})^{2j_{2}-1}(t-y_{i})
\prod^{k}_{i<j}(y_{i}-y_{j})^{-2}
\label{L72}
\eea
The contours in (\ref{L71}),(\ref{L72}) go one below the other,
all in the range $0<z<1$.

The following results have been found for these integrals:
\bea
&&J_{M}(t)=J_{M}(0)F_{M}(t)
\label{L73}\\
&&J_{M}(0)=k!(-1)^{k}
\prod^{k}_{1}\frac{\Gamma(i)
\Gamma(2+2j_{3}+i)}{\Gamma(2+2j_{1}-i)\Gamma(2+2j_{2}-i)}
\label{L74}\\
&&J_{L}(t)=J_{L}(0)F_{L}(t)
\label{L75}\\
&&J_{M}(0)=\frac{(-1)^{k}}{\Gamma(-1)}^{k}
\prod^{k}_{1}\frac{\Gamma(1+2j_{1}-i)
\Gamma(1+2j_{2}-i)}{\Gamma(1+i)\Gamma(1+2j_{3}+i)}
\label{L76}\\
&&F_{M}(t)=F_{L}(t)=
{\cal F}(1+j_{1}-j_{2}+j_{3},-k;2+2j_{3};t)
\label{L77}
\eea
In (\ref{L77}) ${\cal F}(\alpha,\beta;\gamma;t)$ is
the hypergeometric function -- a polynomial, in fact,
of order $k$, as $\beta=-k$, and $k$ is always assumed
to be integer. This form of $F_{M}(t)$, $F_{L}(t)$
is an assumption, verified by calculating the first
two and the last coefficients of the expansions
of $J_{M}(t)$, $J_{L}(t)$ in powers of $t$.

Using the parametrization
\beq
a=-2j_{2}-1, \quad b=2j_{3}
\label{L78}
\eeq
the function (\ref{L77}) could be presented in the following
form:
\beq
{\cal F}(t)=(-1)^{k}k!\frac{\Gamma(2+b)}{\Gamma(2+k+b)}
P^{(a,b+1)}_{k}(2t-1)
\label{L79}
\eeq
where $P^{(\alpha,\beta)}_{k}$ is the Jacobi polynomial.

On account of (\ref{L69}-\ref{L79}) the integral (\ref{L65})
could be transformed as
\bea
&&I=\frac{\sin\pi(-2j_{2}-1)}{\pi}J_{M}(0)J_{L}(0)
\int^{\infty}_{1}du(u)^{-(2j_{1}+1)}(u-1)^{-(2j_{2}+1)}
\times (u)^{2k}({\cal F}(\frac{1}{u}))^{2}
\nn\\
&&=\frac{\sin\pi(-2j_{2}-1)}{\pi}J_{M}(0)J_{L}(0)
\int^{1}_{0}du(u)^{2j_{3}}(1-u)^{-2j_{2}-1}
({\cal F}(u))^{2}
\nn\\
&&=\frac{\sin\pi(-2j_{2}-1)}{\pi}J_{M}(0)J_{L}(0)
(k!\frac{\Gamma(2+b)}{\Gamma(2+k+b)})^{2}
\nn\\
&&\times \int^{1}_{0}du(u)^{b}(1-u)^{a}
(P^{(a,b+1)}_{k}(2u-1))^{2}
\label{L80}
\eea
The parameters $a$, $b$ are defined in (\ref{L78}).
The last integral could be found in the tables
and is given by
\beq
\frac{\Gamma(2+b+k)\Gamma(1+a+k)}{k!(b+1)\Gamma(2+a+b+k)}
\label{L81}
\eeq
Substituting the expressions for $J_{M}(0)$, $J_{L}(0)$
in (\ref{L74}), (\ref{L76}), and the expression (\ref{L81})
for the integral in (\ref{L80}), we find:
\beq
I=\frac{1}{(\Gamma(-1))^{k}}(j_{1}+j_{2}-j_{3})!
\frac{\Gamma(1+2j_{3})}{\Gamma(1+2j_{1})\Gamma(1+2j_{2})}
\label{L82}
\eeq

This is the value of the coefficient in the operator
product expansion (\ref{L63}) and (\ref{L61}),
the factor $(1/\Gamma(-1))^{k}$ is to be absorbed
in the renormalization of $(\mu)^{k}$ which has been dropped
in the above expressions. On general grounds this coefficient
has to be of the form:
\beq
(^{j_{1}\quad\,\,\,\,\,\, j_{2}\quad\,\,\,\,\,\, j_{3}}_{j_{1}\quad
j_{3}-j_{1}\quad -j_{3}})
d_{j_{1},j_{2},j_{3}}'
\label{L83}
\eeq
like that in Section 2.
In a similar way, by matching (\ref{L83}) with (\ref{L82}),
the coefficient $d_{j_{1},j_{2},j_{3}}'$ is found to be:
\bea
&&d_{j_{1},j_{2},j_{3}}'=(\mu_{eff})^{j_{1}+j_{2}-j_{3}}
\times\frac{(j_{1}+j_{2}-j_{3})!(j_{1}-j_{2}+j_{3})!(-j_{1}
+j_{2}+j_{3})!}{\Delta^{1/2}_{j_{1},j_{2},j_{3}}[(2j_{1})!
(2j_{2})!(2j_{3})!]^{1/2}}
\nn\\
&&\quad\quad\quad\quad\quad\quad\quad\quad\quad\quad\quad
\times
\frac{(2j_{3})!}{(2j_{1})!(2j_{2})!}
\label{L84}
\eea
Here we have supplied back the factor
$(\mu_{eff})^{k}$. The first and the last factors
in (\ref{L84}) could be absorbed into the normalization
of the operators, and we get the symmetric coefficients
$d^{(B)}_{j_{1},j_{2},j_{3}}$ in (\ref{L39}).

\section{Conclusions and discussion}

One conclusion of the calculation is that the deformed
operator algebra of currents is different from
${\cal T}(\mu)$, the enveloping algebra of $su(2)$ for
a fixed value $\mu$ of the quadratic Casimir operator,
which has been discussed in a number of recent papers.

The relation of the operators $O_{j,m}$ to $T_{j,m}$
by the operator $\delta$ in (\ref{L29}) might be of interest
on its own, apart from the purpose of the calculation
in this paper.

The relation by the linear transformation (\ref{L11})
of the minimal, $C<1$, and $C=1$ theories suggests the use
of extra screening operators, of Liouville--matter mixed
type. They would appear in particular in the representation
of correlation functions of fractional $j$ operators
(\ref{L14}),(\ref{L15}), in the context of $C=1$ theory.
The correlation function of this type are related
to the unsolved yet problem of calculating the general
multipoint functions in $C<1$ theory,-- to begin with
the four--point ones, still on a sphere. Remarks to this
problem are given in \cite{a22}.
We hope to make some progress on this in the future.

\noindent{\large Acknowledgements}

I am grateful to the colleagues at CERN and at LPTHE,
Jussieu, Paris, for the hospitality and stimulating
discussions. I have benefited in particular from
discussions with E. Kiritsis and with P. Bouwknegt.

\newpage


\begin{thebibliography}{10}

\bibitem{a1} D.~J.~Gross, I.~R.~Klebanov and M.~J.~Newman,
{\it Nucl.Phys.}~{\bf B350} (1991) 621;\\
D.~J.~Gross, I.~R.~Klebanov,
{\it Nucl.Phys.}~{\bf B352} (1991) 671

\bibitem{a2} B.~Lian and G.~Zuckerman, {\it Phys.Lett.}~
{\bf 254B} (1991) 417; {\it Phys.Lett.}~{\bf 266B} (1991)
21

\bibitem{a3} A.M.Polyakov, {\it Mod.Phys.Lett.}~{\bf A6}
(1991) 635

\bibitem{a4} C.~Imbimbo, S.~Mahapatra, and S.~Mukhi,
Preprint GEF-TH 8/91, TIFR/TH/91-27, May 1991

\bibitem{a5} P.~Bouwknegt, J.~McCarthy, and
K.~Pilch, Preprint CERN-TH.6162/91, July 1991

\bibitem{a6} S.~Govindarajan, T.~Jayaraman, V.~John and
P.~Majumdar, Preprint IMSc -91/40, December 1991

\bibitem{a7} N.~Chair, V.~K.~Dobrev and H.~Kanno,
Preprint IC/92/17, January 1992

\bibitem{a8} E.~Witten, Preprint SLAC-PUB-IASSNS-HEP-91/51,
August, 1991

\bibitem{a9} I.~R.~Klebanov and A.~M.~Polyakov,
{\it Mod.Phys.Lett.}~{\bf A6} (1991) 3273

\bibitem{a10} J.~Avan and A.~Jevicki, {\it Phys.Lett.}~
{\bf 266B} (1991) 35; Preprints BROWN-HET-824 and
BROWN-HET-839;\\
D.~Mimic, J.~Polchinski and Z.~Yang, Preprint UTTG-16-91;\\
G.~Moore and N.~Seiberg, Preprint RU-91-29, YCTP-P19-91;\\
S.~Das, A.~Dhar, G.~Mandal and S.~Wadia, Preprints
IASSNS-HEP-91/52 and 91/72

\bibitem{a11} I.~R.~Klebanov, {\it Mod.Phys.Lett.}~
{\bf A7} (1992) 723

\bibitem{a12} E.~Verlinde, Preprint IASSNS-HEP-92/5,
February 1992

\bibitem{a13} E.~Witten and B.~Zwiebach,
Preprint SLAC-PUB-IASSNS-HEP-92/2, MIT-CTP-2057,
January 1992

\bibitem{a131} M.~Bershadsky and D.~Kutasov,
Preprint PUPT-1315, HUTP-92/A016, April 1992

\bibitem{a132} I.~R.~Klebanov and A.~Pasquinucci,
Preprint PUPT-1313, April 1992

\bibitem{a14} M.~Li, Preprint UCSBTH-91-47, October 1991

\bibitem{a15} S.~Kachru, Preprint PUPT-1305, January 1992

\bibitem{a16} J.~L.~F.~Barb\'on, Preprint CERN-TH.6379/92,
FTUAM-92-02, January 1992

\bibitem{a17} Vl.~S.~Dotsenko, Preprint PAR-LPTHE 92-4,
January 1992

\bibitem{a18} F.~David, {\it Mod.Phys.Lett.}~{\bf A3} (1988) 1651\\
J.~Distler and H.~Kawai, {\it Nucl.Phys.}~{\bf B321}
(1989) 509

\bibitem{a19} D.~Friedan, E.~Martinec and S.~Shenker,
{\it Nucl.Phys.}~{\bf B271} (1986) 93

\bibitem{a20} Vl.~S.~Dotsenko and V.~A.~Fateev,
{\it Nucl.Phys.}~{\bf B251} (1985) 691

\bibitem{a21} Vl.~S.~Dotsenko, {\it Adv.Stud. in Pure Math.}~{\bf 16}
(1988)123

\bibitem{a22} Vl.~S.~Dotsenko, Preprint PAR-LPTHE 91-52,
to be published in Proceedings of Carg\`ese Summer School,
July 15 -- July 27 1991



\end{thebibliography}
\end{document}